\documentclass[conference]{IEEEtran}
\usepackage{amsmath,amsthm,amssymb}
\usepackage{mathrsfs,mathtools,mathabx}
\usepackage{bm}
\usepackage{graphicx}

\usepackage{dsfont,enumitem,stackrel}
\usepackage{color}
\usepackage{array}
\usepackage{cases}
\usepackage{algorithm,algpseudocode}
\usepackage{algpseudocode}
\usepackage{pgfplots}
\usepackage{subfigure}
\pgfplotsset{compat=newest}
\usepackage{accents}
\usepackage{float}
\usepackage{cite}
\usepackage{xparse}
\usepackage{xcolor}
\usepackage{balance}
\usepackage{tikz}
\usetikzlibrary{arrows,decorations.pathreplacing,patterns}

\usepackage[hyphens]{url}
\usepackage{hyperref}
\usepackage[hyphenbreaks]{breakurl}

\usepackage{lipsum}
\usepackage{booktabs} 

\definecolor{chestnut}{rgb}{0.8, 0.36, 0.36}
\definecolor{airforceblue}{rgb}{0.36, 0.54, 0.66}
\definecolor{cadmiumorange}{rgb}{0.93, 0.53, 0.18}
\definecolor{bleudefrance}{rgb}{0.19, 0.55, 0.91}
\definecolor{carolinablue}{rgb}{0.6, 0.73, 0.89}
\definecolor{blue(ncs)}{rgb}{0.0, 0.53, 0.74}
\definecolor{dodgerblue}{rgb}{0.12, 0.56, 1.0}
\definecolor{cssgreen}{rgb}{0.0, 0.5, 0.0}
\definecolor{cadmiumgreen}{rgb}{0.0, 0.42, 0.24}
\definecolor{cadmiumorange}{rgb}{0.93, 0.53, 0.18}
\definecolor{amaranth}{rgb}{0.9, 0.17, 0.31}
\definecolor{bluegray}{rgb}{0.4, 0.6, 0.8}
\definecolor{cadmiumgreen}{rgb}{0.0, 0.42, 0.24}

\graphicspath{{./Figures/}} 

\newtheorem{prop}{Proposition}

\newtheorem{remark}{Remark}

\theoremstyle{definition}

\newcommand{\Comp}{ChannelComp~}

\begin{document}
%
\IEEEoverridecommandlockouts
\title{ Computing Functions Over-the-Air 

Using Digital Modulations}


%

\author{\IEEEauthorblockN{Saeed Razavikia$^\dagger$, José Mairton Barros da Silva Jr$^{*,\dagger}$, and Carlo Fischione$^\dagger$} 
\IEEEauthorblockA{$^\dagger$School of Electrical Engineering and Computer Science, KTH Royal Institute of Technology, Stockholm, Sweden\\
$^*$Department of Electrical and Computer Engineering, Princeton University, New Jersey, USA\\
Email: \{sraz, jmbdsj, carlofi\}@kth.se}
	\thanks{S. Razavikia was supported by the Wallenberg AI, Autonomous Systems and Software Program (WASP).}
 	\thanks{Jose Mairton Barros da Silva Jr. was supported by the European Union’s Horizon Europe research and innovation program under the Marie Skłodowska-Curie project FLASH, with grant agreement No 101067652.} 
    \thanks{The EU FLASH project, the Digital Futures project DEMOCRITUS, and the Swedish Research Council Project MALEN partially supported this work.}
}

\maketitle

\begin{abstract}	
Over-the-air computation  (AirComp) is a known technique in which wireless devices transmit values by analog amplitude modulation so that a function of these values is computed over the communication channel at a common receiver. The physical reason is the superposition properties of the electromagnetic waves, which naturally return sums of analog values.  Consequently, the applications of AirComp are almost entirely restricted to analog communication systems. However, the use of digital communications for over-the-air computations would have several benefits, such as error correction, synchronization, acquisition of channel state information, and easier adoption by current digital communication systems.
Nevertheless, a common belief is that digital modulations are generally unfeasible for computation tasks because the overlapping of digitally modulated signals returns signals that seem to be meaningless for these tasks. 
This paper breaks through such a belief and  proposes a fundamentally new  computing method, named ChannelComp, for performing over-the-air computations by any digital modulation. In particular, we propose digital modulation formats that allow us to compute a wider class of functions than AirComp can compute, and we propose a feasibility optimization problem that ascertains the optimal digital modulation for computing functions over-the-air. The simulation results verify the superior performance of ChannelComp in comparison to AirComp,  particularly for the product functions, with around $10$ dB improvement of the computation error.  
\end{abstract}
\begin{IEEEkeywords}
  Digital communication, modulation, nomographic functions, over-the-air computation, symmetric function.   
\end{IEEEkeywords}

\section{Introduction}

The increasing number of internet of things (IoT) devices and applications using machine learning (ML) techniques require extensive connectivity. This implies scaling up radio and computing resources, and potentially saturating the capacity of current systems \cite{tataria20216g,pham2022aerial}. Consequently, to better support such compute-intensive applications, the over-the-air computation (AirComp) method has emerged as a promising concept where collecting data or values and performing computations over them  simultaneously occurs at the edge network  \cite{zhu2019broadband,goldenbaum2013harnessing,nazer2007computation}. 

The AirComp method executes/performs  the computation of mathematical functions of the devices' data by leveraging the waveform superposition property of wireless communication channels. Unlike the standard transmit-then-compute scheme, AirComp brings a high-rate communication scheme for multiple access channels (MACs) by harnessing interference to help functional computations. AirComp provides ultra-fast wireless data aggregation in IoT networks with high spectrum efficiency.
Compared to the standard transmit-then-compute schemes, AirComp can dramatically reduce the required communication resources (power, bandwidth, channel usage, etc.), particularly in distributed learning, where it has also attracted growing attention in federated edge learning  \cite{yang2020federated,amiri2020federated,zhu2019broadband,hellstrom2022wireless}.

However, AirComp entirely relies on analog communication, which is difficult for reliable communications due to channel implications \cite{csahin2022over}. In addition, AirComp requires analog hardware systems for deploying analog modulation, which is a drawback due to the limited number of current wireless devices that support analog modulations. It would be more advantageous to use digital modulations, given their good properties in terms of channel correction, source and channel coding, and widespread use. This, however, is believed to be extremely difficult due to that the overlapping of digitally modulated signals returns, in general, incomprehensible signals for function computations\cite{zhu2019broadband,wang2022over,kaibin2022digital}. 

\input{fig/Fig_System_model.tex}
As an attempt towards digital AirComp, recently, the authors in \cite{zhu2020one} have considered a machine learning set-up and have proposed a broadband one-bit for aggregation over-the-air (OBDA) based on a majority vote for solving the signSGD problem \cite{bernstein2018signsgd} while using binary phase shift keying (BPSK) modulations. Further, OBDA extensions to other modulations, e.g., frequency-shift keying (FSK) \cite{csahin2021distributed}, and an asynchronous OFDM-based version of OBDA have been proposed in \cite{zhao2021broadband}.  AirComp's non-coherent communication solution for single and multi-cell using pulse-position modulation and FSK have been studied in \cite{sahin2021over,csahin2021distributed,hassan2022multi}. All these OBDA studies are limited to a specific function (sign function) or specific ML training procedure (signSGD problem). Consequently, unlike function computations in the AirComp method, existing attempts to use digital modulations with AirComp  cannot compute larger classes of functions beyond nomographic functions \cite{goldenbaum2014nomographic} and are unsuitable for general digital modulations beyond the simple BPSK or FSK.

In this paper, we propose a fundamentally new digital channel computation method, termed \textit{ChannelComp}, for computing functions over MAC by any digital modulation format and for a class of functions more general than the one that can be handled by AirComp. Specifically, we  consider the problem of computing a K-variate function $f(x_1, \ldots, x_K): \mathbb{R}^{K}\mapsto \mathbb{R}$, where $x_k \in \mathbb{R}$ for $k =1, \ldots, K,$ belongs to node $k$ of a network with $K$ nodes. The nodes use digital modulations to transmit the values $x_k$ over a MAC to a server that needs to compute the function $f$ of the nodes' values. We establish the conditions for  computing a class of functions more general than the class that AirComp can handle. For a given function, these conditions lead to optimization problems whose solutions determine the parameters of the digital modulation resulting in a correct computation over-the-air. Specifically, we propose a feasibility optimization problem to obtain the parameters of the used digital modulation. Such a feasibility problem is NP-hard, and to overcome such complexity, we develop a convex relaxation that can be solved using traditional solvers, such as CVX \cite{grant2014cvx}. 

\Comp can  compute functions for a finite number of its input domain, which is the typical case of digital communication systems because they only handle quantized values. Further, \Comp provides a wireless aggregation communication system at least as fast as AirComp. This is because  \Comp adapts the parameters of the digital modulation format such that the receiver computes the desired function without re-transmissions or error corrections,  which leads to a low latency computation over-the-air. In the numerical experiments we present in this paper, \Comp outperforms AirComp in terms of computation error for various important functions while consuming the same communication resources. For example, for computing the product function, \Comp obtains a $10$ dB performance improvement compared to AirComp in terms of the normalized mean square error, without needing to use analog modulations but only relying on currently widespread digital modulations.

The rest of the paper is organized as follows: in Section \ref{sec:model}, we explain the system model and present our proposed ChannelComp method. Then, we characterize how to select the digital modulation formats for computing the desired function over the MAC in Section \ref{sec:fun}.  In Section \ref{sec:Num}, we present  the numerical experiments and the performance comparison between ChannelComp and AirComp. Finally, we conclude the paper in Section \ref{sec:conclusion}.
\vspace{-4pt}
\subsection{Notation}
Bold lower-case letters $\bm{x}$ are used to indicate vector quantities, and bold upper-case $\bm{X}$ to denote matrices. We use $\vec{x}$ to show modulated band-pass signals.
The transpose and Hermitian of a matrix $\bm{X}$ are represented by $\bm{X}^{\mathsf{T}}$ and $\bm{X}^{\mathsf{H}}$, respectively. We use $\otimes$ to represent the Kronecker product.  We denote the range set of the function $f$ by $R_f$, and its cardinality by $|R_f|$.

For an integer $N$, $[N]$ stands for $\{1,2,\ldots, N\}$. A finite integer field of size $q$ with a subset of the integers number is represented by $\mathbb{F}_q\subset \mathbb{Z}$. We use $\bm{X}\succeq \bm{0}$ to show that $\bm{X}$ is a positive semidefinite matrix. Finally, for two arbitrary matrices $\bm{A}, \bm{B}$, $\langle \bm{A}, \bm{B}\rangle$ represents the inner product which is equal to the trace of $\bm{B}^{\mathsf{H}}\bm{A}$.

\section{System Model and Method Formulation}\label{sec:model}
We consider a system model similar to AirComp's system model, except for the encoders, which use digital modulations instead of amplitude analog modulations.  For ease of explanation, we consider narrow-band commutations. The extension to broadband communication using the orthogonal frequency division multiple access (OFDMA) method is straightforward~\footnote{We can add a conventional  \textit{orthogonal frequency division multiplexing} module \cite{zhu2020one} for all nodes, right after encoder module in Fig. \ref{fig:ComSystem}, and thus we can perform multiple computations per frequency carrier.}. 

Consider a communication network with $K$ nodes and a server as a computation point (CP). In such a network, the nodes communicate via a shared communication channel with the CP. We aim to compute a function  $f(x_1,x_2,\ldots,x_K)$ at the CP via the communication channel, where $x_k\in \mathbb{R}$ denotes one of the function's input values, namely the value that node $k$ generates. We consider a class of functions more general than the class of functions that AirComp can handle (nomographic functions), that is, we consider the class of symmetric functions. \hspace{-0.5pt} Recall that a function $f$ is symmetric if    
\begin{align}
  \label{eq:sym}
  f(x_1,\ldots,x_K) = f(\pi(x_1),\ldots,\pi(x_K)),  
\end{align}
for all possible permutations by $\pi: \{1,\ldots,K\}\mapsto \{1,\ldots,K\}$. For example, $\sum_{k=1}^Kx_k$ or $\prod_{k=1}^Kx_k$ are symmetric functions. Note that the class of nomographic functions assumes the form  $f =\psi \big(\sum_{k=1}^K\varphi_k(x_k)\big)$, where $\psi,\varphi_1,\ldots, \varphi_K: \mathbb{R}\mapsto \mathbb{R}$, and thus we  can use $w_k:= \varphi_k(x_k)$ for $k=1,\ldots, K$, and turn any nomographic function $f$ into a symmetric function, i.e., $f(w_1,\ldots,w_k) = \psi(\sum_{k=1}^Kw_k)$. Thus, similar to optimization theory, where changes in variables do not change the optimum's computation, we can work with symmetric functions and be more general than nomographic functions.

Because we are using digital communications, node $k$ quantizes its value $x_k$ to $\tilde{x}_k \in \mathbb{F}_q$, where $q$ stands for the number of quantization levels. Then, node $k$ sends $\tilde{x}_k$ using a digital modulation format over the MAC to compute the function $f$ at the CP. In particular, the value $\tilde{x}_k\in \mathbb{F}_q$  is mapped into the digitally modulated signal $\vec{x}_k$  using the digital encoder $\mathscr{E}(\cdot)$.  Afterward, all the nodes transmit simultaneously\footnote{It is assumed that we have perfect synchronization among all the nodes and the CP. However, the existing techniques of analog Aircomp for solving imperfect synchronization, e.g.,  \cite{saeed2022BlindFed} can be applied to our system model due to the similarity of the proposed system model with the AirComp model. } the modulated values $\vec{x}_k$'s over the MAC. The CP receives the summation $\vec{y}$ through the superposition nature of waves, i.e.,   
\begin{align}
    \label{eq:Aggnoise}
      \vec{y} = \sum\nolimits_{k=1}^{K}h_k p_k\vec{x}_k + \vec{z},
\end{align}
where $h_k$ denotes the channel attenuation between node $k$ and the CP, $p_k$ is the transmit power used by node $k$, and $\vec{z}$ is the additive white Gaussian noise (AWGN) with zero-mean and variance $\sigma_z^2$.  Then, the method that ChannelComp proposes in this paper is to compute the function $f$ just by looking at the received signal $\vec{y}$.

The general belief is that the computation of the function by looking at $\vec{y}$ is very difficult or impossible since $\vec{y}$ is the superposition of digital signals, which in general is not a linear combination of the transmit values $x_k$, in contrast to what would happen if we used analog modulations. However, even though $\vec{y}$ does not give a sum of the $x_k$'s, it is still useful to perform a computation. The intuitive reason is that the transmitted $\vec{x}_k$'s are constructed from finite constellation points, and the received signal $\vec{y}$, in the absence of AWGN,  may have a reshaped constellation of a finite number of points. Therefore, we can use a mapping (look-up table) function $\mathscr{T}(\vec{y})$ on the received constellation diagram of signal $\vec{y}$ to obtain the correct output of the function $f$. The overall system model is depicted in Fig. \ref{fig:ComSystem}.

In the following section, we explain in greater detail the intuitive idea we have given above, and we propose a structure for the  encoder and the decoder of ChannelComp.

\section{Communication Architecture of \Comp }\label{sec:fun}

In this section, we describe how to design the encoder and decoder of ChannelComp. We follow the power control universally adopted in the AirComp literature \cite{cao2019optimal} and select the transmit power as the inverse of the channel, i.e., $p_k = h_k^{*}/|h_k|^{2}$. Hence, Eq.~~\eqref{eq:Aggnoise} becomes
\begin{align}
    \label{eq:noisefree}
      \vec{y} = \sum\nolimits_{k=1}^{K}\vec{x}_k + \vec{z}.
\end{align}
From now on, without loss of generality, we use Eq. \eqref{eq:noisefree}.

In the following Subsection \ref{sec:Encoder}, we introduce the working principle of the ChannelComp encoder to a specific case of two nodes and BPSK. In Section \ref{sec:encoderK}, we introduce the general design of the encoder for a $K$-nodes system, and in Section \ref{sec:Reciever}, we introduce how to design the decoder.

\subsection{ Encoder for $2$ nodes, BPSK, and noiseless MAC}\label{sec:Encoder}
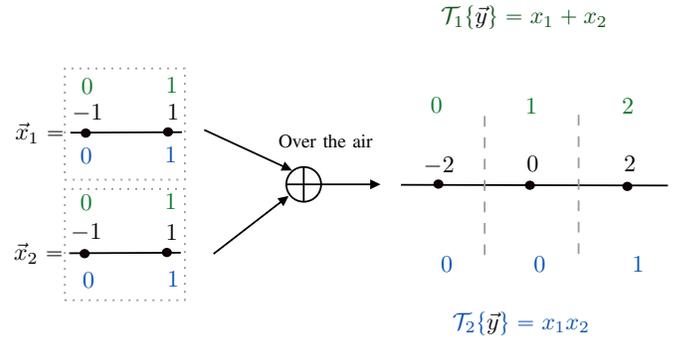
\begin{figure}
    \centering
    
    \scalebox{0.9}{ 

\tikzset{every picture/.style={line width=0.75pt}} 

\begin{tikzpicture}[x=0.75pt,y=0.75pt,yscale=-0.65,xscale=0.65]

\draw   (259,190) .. controls (259,181.72) and (265.72,175) .. (274,175) .. controls (282.28,175) and (289,181.72) .. (289,190) .. controls (289,198.28) and (282.28,205) .. (274,205) .. controls (265.72,205) and (259,198.28) .. (259,190) -- cycle ; \draw   (259,190) -- (289,190) ; \draw   (274,175) -- (274,205) ;
\draw    (188.25,143) -- (261.28,176.74) ;
\draw [shift={(264,178)}, rotate = 204.8] [fill={rgb, 255:red, 0; green, 0; blue, 0 }  ][line width=0.08]  [draw opacity=0] (8.93,-4.29) -- (0,0) -- (8.93,4.29) -- cycle    ;
\draw    (196.25,250) -- (230.27,223.73) -- (258.63,201.83) ;
\draw [shift={(261,200)}, rotate = 142.32] [fill={rgb, 255:red, 0; green, 0; blue, 0 }  ][line width=0.08]  [draw opacity=0] (8.93,-4.29) -- (0,0) -- (8.93,4.29) -- cycle    ;
\draw    (288.25,190) -- (337.25,190) ;
\draw [shift={(340.25,190)}, rotate = 180] [fill={rgb, 255:red, 0; green, 0; blue, 0 }  ][line width=0.08]  [draw opacity=0] (8.93,-4.29) -- (0,0) -- (8.93,4.29) -- cycle    ;
\draw    (71.83,249.4) -- (168,249) ;
\draw  [fill={rgb, 255:red, 20; green, 1; blue, 1 }  ,fill opacity=1 ] (152.51,248.76) .. controls (152.51,247) and (154.08,245.57) .. (156.03,245.57) .. controls (157.97,245.57) and (159.55,247) .. (159.55,248.76) .. controls (159.55,250.53) and (157.97,251.96) .. (156.03,251.96) .. controls (154.08,251.96) and (152.51,250.53) .. (152.51,248.76) -- cycle ;
\draw  [fill={rgb, 255:red, 20; green, 1; blue, 1 }  ,fill opacity=1 ] (81.57,249.76) .. controls (81.57,248) and (83.14,246.57) .. (85.09,246.57) .. controls (87.03,246.57) and (88.61,248) .. (88.61,249.76) .. controls (88.61,251.53) and (87.03,252.96) .. (85.09,252.96) .. controls (83.14,252.96) and (81.57,251.53) .. (81.57,249.76) -- cycle ;
\draw    (357.83,190.4) -- (588,191) ;
\draw  [fill={rgb, 255:red, 20; green, 1; blue, 1 }  ,fill opacity=1 ] (465.51,190.76) .. controls (465.51,189) and (467.08,187.57) .. (469.03,187.57) .. controls (470.97,187.57) and (472.55,189) .. (472.55,190.76) .. controls (472.55,192.53) and (470.97,193.96) .. (469.03,193.96) .. controls (467.08,193.96) and (465.51,192.53) .. (465.51,190.76) -- cycle ;
\draw  [fill={rgb, 255:red, 20; green, 1; blue, 1 }  ,fill opacity=1 ] (386.57,189.76) .. controls (386.57,188) and (388.14,186.57) .. (390.09,186.57) .. controls (392.03,186.57) and (393.61,188) .. (393.61,189.76) .. controls (393.61,191.53) and (392.03,192.96) .. (390.09,192.96) .. controls (388.14,192.96) and (386.57,191.53) .. (386.57,189.76) -- cycle ;
\draw  [fill={rgb, 255:red, 20; green, 1; blue, 1 }  ,fill opacity=1 ] (549.51,191.76) .. controls (549.51,190) and (551.08,188.57) .. (553.03,188.57) .. controls (554.97,188.57) and (556.55,190) .. (556.55,191.76) .. controls (556.55,193.53) and (554.97,194.96) .. (553.03,194.96) .. controls (551.08,194.96) and (549.51,193.53) .. (549.51,191.76) -- cycle ;
\draw [color={rgb, 255:red, 156; green, 156; blue, 156 }  ,draw opacity=1 ] [dash pattern={on 4.5pt off 4.5pt}]  (430,131) -- (430,252) ;
\draw [color={rgb, 255:red, 156; green, 156; blue, 156 }  ,draw opacity=1 ] [dash pattern={on 4.5pt off 4.5pt}]  (511,130) -- (511,251) ;
\draw    (72.83,145.4) -- (169,145) ;
\draw  [fill={rgb, 255:red, 20; green, 1; blue, 1 }  ,fill opacity=1 ] (153.51,144.76) .. controls (153.51,143) and (155.08,141.57) .. (157.03,141.57) .. controls (158.97,141.57) and (160.55,143) .. (160.55,144.76) .. controls (160.55,146.53) and (158.97,147.96) .. (157.03,147.96) .. controls (155.08,147.96) and (153.51,146.53) .. (153.51,144.76) -- cycle ;
\draw  [fill={rgb, 255:red, 20; green, 1; blue, 1 }  ,fill opacity=1 ] (82.57,145.76) .. controls (82.57,144) and (84.14,142.57) .. (86.09,142.57) .. controls (88.03,142.57) and (89.61,144) .. (89.61,145.76) .. controls (89.61,147.53) and (88.03,148.96) .. (86.09,148.96) .. controls (84.14,148.96) and (82.57,147.53) .. (82.57,145.76) -- cycle ;
\draw  [color={rgb, 255:red, 155; green, 155; blue, 155 }  ,draw opacity=1 ][dash pattern={on 0.84pt off 2.51pt}][line width=0.75]  (68,90) -- (171.5,90) -- (171.5,186) -- (68,186) -- cycle ;
\draw  [color={rgb, 255:red, 155; green, 155; blue, 155 }  ,draw opacity=1 ][dash pattern={on 0.84pt off 2.51pt}][line width=0.75]  (68,194) -- (171.5,194) -- (171.5,290) -- (68,290) -- cycle ;

\draw (250,145) node [anchor=north west][inner sep=0.75pt]  [font=\footnotesize] [align=left] {Over the air};
\draw (71.21,221.9) node [anchor=north west][inner sep=0.75pt]   [align=left] {$\displaystyle -1$};
\draw (152.93,222.18) node [anchor=north west][inner sep=0.75pt]   [align=left] {$\displaystyle 1$};
\draw (375.21,163.9) node [anchor=north west][inner sep=0.75pt]   [align=left] {$\displaystyle -2$};
\draw (463.93,163.18) node [anchor=north west][inner sep=0.75pt]   [align=left] {$\displaystyle 0$};
\draw (547.93,163.18) node [anchor=north west][inner sep=0.75pt]   [align=left] {$\displaystyle 2$};
\draw (381,112.4) node [anchor=north west][inner sep=0.75pt]  [color={rgb, 255:red, 8; green, 121; blue, 43 }  ,opacity=1 ]  {$0$};
\draw (463,113.4) node [anchor=north west][inner sep=0.75pt]  [color={rgb, 255:red, 8; green, 121; blue, 43 }  ,opacity=1 ]  {$1$};
\draw (546,113.4) node [anchor=north west][inner sep=0.75pt]  [color={rgb, 255:red, 8; green, 121; blue, 43 }  ,opacity=1 ]  {$2$};
\draw (390,33.4) node [anchor=north west][inner sep=0.75pt]  [color={rgb, 255:red, 15; green, 83; blue, 26 }  ,opacity=1 ]  {$\mathcal{T}_1\{ {\color{black}\vec{y}}\} =x_{1} +x_{2}$};
\draw (22,238.4) node [anchor=north west][inner sep=0.75pt]    {$\Vec{x}_{2} =$};
\draw (72.21,118.9) node [anchor=north west][inner sep=0.75pt]   [align=left] {$\displaystyle -1$};
\draw (153.93,118.18) node [anchor=north west][inner sep=0.75pt]   [align=left] {$\displaystyle 1$};
\draw (23,134.4) node [anchor=north west][inner sep=0.75pt]    {$\Vec{x}_{1} =$};
\draw (400,297.4) node [anchor=north west][inner sep=0.75pt]  [color={rgb, 255:red, 8; green, 80; blue, 177 }  ,opacity=1 ]  {$\mathcal{T}_2\{{\color{black}\vec{y}}\} =x_{1} x_{2}$};
\draw (555,249.4) node [anchor=north west][inner sep=0.75pt]  [color={rgb, 255:red, 8; green, 80; blue, 177 }  ,opacity=1 ]  {$1$};
\draw (470,249.4) node [anchor=north west][inner sep=0.75pt]  [color={rgb, 255:red, 8; green, 80; blue, 177 }  ,opacity=1 ]  {$0$};
\draw (390,249.4) node [anchor=north west][inner sep=0.75pt]  [color={rgb, 255:red, 8; green, 80; blue, 177 }  ,opacity=1 ]  {$0$};
\draw (80,96.4) node [anchor=north west][inner sep=0.75pt]  [color={rgb, 255:red, 8; green, 121; blue, 43 }  ,opacity=1 ]  {$0$};
\draw (153,95.4) node [anchor=north west][inner sep=0.75pt]  [color={rgb, 255:red, 8; green, 121; blue, 43 }  ,opacity=1 ]  {$1$};
\draw (79,196.4) node [anchor=north west][inner sep=0.75pt]  [color={rgb, 255:red, 8; green, 121; blue, 43 }  ,opacity=1 ]  {$0$};
\draw (152,195.4) node [anchor=north west][inner sep=0.75pt]  [color={rgb, 255:red, 8; green, 121; blue, 43 }  ,opacity=1 ]  {$1$};
\draw (79,156.4) node [anchor=north west][inner sep=0.75pt]  [color={rgb, 255:red, 8; green, 80; blue, 177 }  ,opacity=1 ]  {$0$};
\draw (152,155.4) node [anchor=north west][inner sep=0.75pt]  [color={rgb, 255:red, 8; green, 80; blue, 177 }  ,opacity=1 ]  {$1$};
\draw (81,263.4) node [anchor=north west][inner sep=0.75pt]  [color={rgb, 255:red, 8; green, 80; blue, 177 }  ,opacity=1 ]  {$0$};
\draw (154,262.4) node [anchor=north west][inner sep=0.75pt]  [color={rgb, 255:red, 8; green, 80; blue, 177 }  ,opacity=1 ]  {$1$};

\end{tikzpicture}
}
\caption{The sum and product computation using the BPSK modulation.  }
    \label{fig:BPSKExample}
\end{figure}
Here we give an intuitive description of the operating principle of our proposed encoder. Assume, in this subsection,  to use  modulations such as BPSK, where encoder $\mathscr{E}(\cdot)$ is 
 \begin{align}
 \label{eq:BPSK}
 \mathscr{E}(\tilde{x}_k) := \begin{cases} A_c , \quad \tilde{x}_k = 1,\\
 				-A_c , \quad \tilde{x}_k = 0, \end{cases}
 \end{align}
where $A_c$ denotes the amplitude of the carrier. 

To illustrate in Fig \ref{fig:BPSKExample}, we depict a simple noiseless scenario for computing the summation $f_1(x_1,x_2)=x_1+x_2$ and the product function (or logic AND function) $f_2(x_1,x_2)=x_1x_2$ with $K=2$ nodes. From this example of Fig. \ref{fig:BPSKExample}, we see that if we increase the modulation order, e.g., QAM $16$, the constellation points given by $\vec{y}$ cannot be uniquely mapped to either summation or product function. Indeed, for different output values of function $f$, the resulting constellation points overlap and merge into  the same point. Thus, computing the value of the function would always be erroneous using QAM $16$ or higher order modulations, even in the noiseless MAC. However, the basic idea is that if we adapt the phase and amplitude of the digital modulation at each node to the desired function to be computed, we can achieve unique constellation points at the receiver, which would allow us to read a look-up table and perform a correct computation. 

Now that we have illustrated the working principle of the encoder, we focus on the general case of many nodes and general digital modulations in the following subsection. 
\subsection{\Comp of $K$ nodes for $K$-variate function}\label{sec:encoderK}

Here, we propose a necessary condition
on the function $f$ in order to compute it uniquely by ChannelComp.

\begin{prop}[Necessary condition]
\label{prop:Necessary}
Let the $K$-variate function $f(x_1,x_2,\ldots,x_K)$ with domain ${D}_f$ where $x_k \in {D}_f$ for $k =1,\ldots,K$ be the symmetric function to compute over the noise-free MAC.  Let each node use the encoder $\mathscr{E}$. Then, function $f$ can be perfectly computed  by the constellation diagram of $\sum_{k=1}^{K}\mathscr{E}({x}_k)$. 
\end{prop}
\begin{proof}
    The proof is by contradiction. Let $K=2$, and $f(x_1,x_2)$ be an  asymmetric function, i.e., $f(a,b)\neq f(b,a)$ where $a, b \in {D}_f$. Then, for a case where $x_1 =a$ and $x_2 = b$, we have $\vec{a}$ and $\vec{b}$ as modulated signal thereof and $\vec{a}+\vec{b}$ would be received by the CP. For the reverse, i.e., $x_1 = b$ and $x_2=a$, the CP also observes $\vec{a}+\vec{b}$. Hence, we have the same constellation point for different values of $f$. However, it is impossible to assign the same vector $\vec{a}+\vec{b}$ to the two different values of $f(a,b)$ and $f(b,a)$ because $f$ is asymmetric.  
\end{proof}

 To see why the condition in Proposition \ref{prop:Necessary} is not sufficient, one can check a simple function $f(x_1,x_2) = x_1x_2$ for the case where the nodes use PAM modulation with two bits $x_k = \{0,1,2,3\}$. For instance, $f(0,2)\!=\!0$ and $f(1,1)\!=\!1$ while their constellation points overlaps and merge into the point $\vec{2}$. 
 
To avoid this overlapping among constellation points, we design the encoder 
such that the computation is error-free.

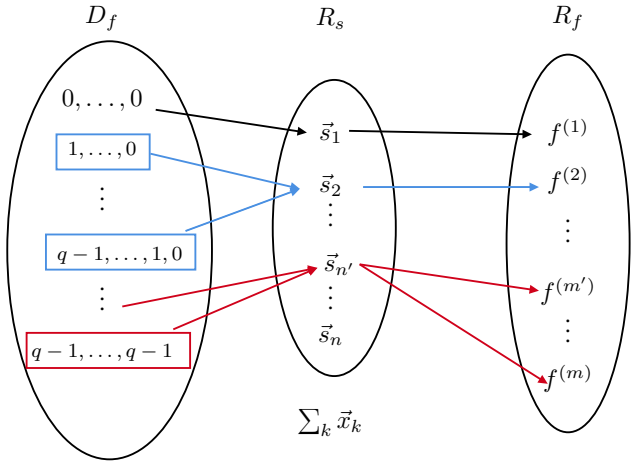
\begin{figure}
    \centering
   \scalebox{0.95}{  

\tikzset{every picture/.style={line width=0.75pt}} 

\begin{tikzpicture}[x=0.7pt,y=0.7pt,yscale=-1,xscale=1]

\draw   (86,152) .. controls (86,86.28) and (112.06,33) .. (144.21,33) .. controls (176.37,33) and (202.43,86.28) .. (202.43,152) .. controls (202.43,217.72) and (176.37,271) .. (144.21,271) .. controls (112.06,271) and (86,217.72) .. (86,152) -- cycle ;
\draw   (370,148.1) .. controls (370,88.73) and (385.67,40.6) .. (405,40.6) .. controls (424.33,40.6) and (440,88.73) .. (440,148.1) .. controls (440,207.47) and (424.33,255.6) .. (405,255.6) .. controls (385.67,255.6) and (370,207.47) .. (370,148.1) -- cycle ;
\draw   (237,139.3) .. controls (237,92.74) and (252.67,55) .. (272,55) .. controls (291.33,55) and (307,92.74) .. (307,139.3) .. controls (307,185.86) and (291.33,223.6) .. (272,223.6) .. controls (252.67,223.6) and (237,185.86) .. (237,139.3) -- cycle ;
\draw    (170.43,72) -- (253.46,84.55) ;
\draw [shift={(256.43,85)}, rotate = 188.6] [fill={rgb, 255:red, 0; green, 0; blue, 0 }  ][line width=0.08]  [draw opacity=0] (5.36,-2.57) -- (0,0) -- (5.36,2.57) -- cycle    ;
\draw [color={rgb, 255:red, 74; green, 144; blue, 226 }  ,draw opacity=1 ]   (167.43,97) -- (249.51,116.31) ;
\draw [shift={(252.43,117)}, rotate = 193.24] [fill={rgb, 255:red, 74; green, 144; blue, 226 }  ,fill opacity=1 ][line width=0.08]  [draw opacity=0] (5.36,-2.57) -- (0,0) -- (5.36,2.57) -- cycle    ;
\draw [color={rgb, 255:red, 74; green, 144; blue, 226 }  ,draw opacity=1 ][fill={rgb, 255:red, 74; green, 144; blue, 226 }  ,fill opacity=1 ]   (187.43,141) -- (249.61,118.04) ;
\draw [shift={(252.43,117)}, rotate = 159.73] [fill={rgb, 255:red, 74; green, 144; blue, 226 }  ,fill opacity=1 ][line width=0.08]  [draw opacity=0] (5.36,-2.57) -- (0,0) -- (5.36,2.57) -- cycle    ;
\draw [color={rgb, 255:red, 74; green, 144; blue, 226 }  ,draw opacity=1 ][fill={rgb, 255:red, 74; green, 144; blue, 226 }  ,fill opacity=1 ]   (288.43,116) -- (385.43,116) ;
\draw [shift={(388.43,116)}, rotate = 180] [fill={rgb, 255:red, 74; green, 144; blue, 226 }  ,fill opacity=1 ][line width=0.08]  [draw opacity=0] (5.36,-2.57) -- (0,0) -- (5.36,2.57) -- cycle    ;
\draw [color={rgb, 255:red, 208; green, 2; blue, 27 }  ,draw opacity=1 ]   (286.43,160) -- (385.46,174.56) ;
\draw [shift={(388.43,175)}, rotate = 188.37] [fill={rgb, 255:red, 208; green, 2; blue, 27 }  ,fill opacity=1 ][line width=0.08]  [draw opacity=0] (5.36,-2.57) -- (0,0) -- (5.36,2.57) -- cycle    ;
\draw [color={rgb, 255:red, 208; green, 2; blue, 27 }  ,draw opacity=1 ]   (286.43,160) -- (390.9,226.39) ;
\draw [shift={(393.43,228)}, rotate = 212.44] [fill={rgb, 255:red, 208; green, 2; blue, 27 }  ,fill opacity=1 ][line width=0.08]  [draw opacity=0] (5.36,-2.57) -- (0,0) -- (5.36,2.57) -- cycle    ;
\draw    (280.43,84) -- (381.43,85.94) ;
\draw [shift={(384.43,86)}, rotate = 181.1] [fill={rgb, 255:red, 0; green, 0; blue, 0 }  ][line width=0.08]  [draw opacity=0] (5.36,-2.57) -- (0,0) -- (5.36,2.57) -- cycle    ;
\draw [color={rgb, 255:red, 208; green, 2; blue, 27 }  ,draw opacity=1 ]   (151.43,184) -- (259.49,162.58) ;
\draw [shift={(262.43,162)}, rotate = 168.79] [fill={rgb, 255:red, 208; green, 2; blue, 27 }  ,fill opacity=1 ][line width=0.08]  [draw opacity=0] (5.36,-2.57) -- (0,0) -- (5.36,2.57) -- cycle    ;
\draw [color={rgb, 255:red, 208; green, 2; blue, 27 }  ,draw opacity=1 ]   (180.43,197) -- (259.67,163.18) ;
\draw [shift={(262.43,162)}, rotate = 156.89] [fill={rgb, 255:red, 208; green, 2; blue, 27 }  ,fill opacity=1 ][line width=0.08]  [draw opacity=0] (5.36,-2.57) -- (0,0) -- (5.36,2.57) -- cycle    ;

\draw (140,66.4) node  {$0,\dotsc ,0$};
\draw (140,118.4) node    {$\vdots $};
\draw (140,175.4) node     {$\vdots $};
\draw (270,85) node    {$\vec{s}_{1}$};
\draw (270,115) node   {$\vec{s}_{2}$};
\draw (275,160) node    {$\vec{s}_{n'}$};
\draw (270,127.4) node     {$\vdots $};
\draw (270,200) node   {$\vec{s}_{n}$};
\draw (270,175) node    {$\vdots $};
\draw (140,20) node    {$D_{f}$};
\draw (270,20) node    {$R_{s}$};
\draw (405,20) node   {$R_{f}$};
\draw (270,250) node     {$\sum _{k}\vec{x}_{k}$};
\draw  [color={rgb, 255:red, 74; green, 144; blue, 226 }  ,draw opacity=1 ]  (114,86) -- (167,86) -- (167,106) -- (114,106) -- cycle  ;
\draw (140,95) node  [font=\footnotesize]  {$1,\dotsc ,0$};
\draw  [color={rgb, 255:red, 74; green, 144; blue, 226 }  ,draw opacity=1 ]  (108,143) -- (193,143) -- (193,163) -- (108,163) -- cycle  ;
\draw (150,155) node   [font=\footnotesize]  {$q-1,\dotsc ,1,0$};
\draw  [color={rgb, 255:red, 208; green, 2; blue, 27 }  ,draw opacity=1 ]  (97,199) -- (190,199) -- (190,219) -- (97,219) -- cycle  ;
\draw (140,210) node  [font=\footnotesize]  {$q-1,\dotsc ,q-1$};
\draw (405,85) node    {$f^{(1)}$};
\draw (405,112) node    {$f^{(2)}$};
\draw (405,225) node     {$f^{(m)}$};
\draw (405,193.4) node     {$\vdots $};
\draw (405,175) node   {$f^{(m')}$};
\draw (405,136.4) node   {$\vdots $};

\end{tikzpicture}
   } 
    \caption{Schematic of the domain $D_f$, the range of the summation $R_s$ returned by a noise-free channel, and desirable function $R_f$ in left, middle, and right of the figure, respectively. Two different points (blue lines) in the domain of function $f$ create a constellation point $\vec{s}_2$ while the function's output for these two points is the same and equal to $f^{(2)}$. However, for $\vec{s}_{n'}$ the corresponding values of the function are different $f^{(m)} \neq f^{(m')}$. Accordingly, we cannot assign  the point $\vec{s}_{n'}$ to these points (red lines), unless we enforce a splitting of $\vec{s}_{n'}$ by a proper selection of the digital modulations. }
    \label{fig:range_vector}
\end{figure}

Let modulation vector $\bm{x}\in \mathbb{C}^{q\times 1}$ be a vector consisting of the values of all possible constellation points resulting from encoder $\mathscr{E}(\cdot)$, e.g., $\bm{x}=[1,-1]$ for BPSK modulation or $\bm{x}=[-3-3j, -1-3j,1-3j\ldots,3+3j] \in \mathbb{C}^{16}$ for QAM $16$. Moreover,  let $R_f$ be the set of all the outputs of the function $f$ with the size of $L = |R_f|$, and the $f^{(i)}\in {R}_f$ denote the $i$-th element, for $i\in [L]$, of the output function $f$ for a certain value of input $x_1,\ldots, x_K$ where all can assume one of $q$ possible values. We further define the complex vector $\vec{\bm{s}}\in \mathbb{C}^{M\times 1}$ with $M:=q^K$, whose element $i$ denotes the induced constellation points from $\sum_{k=1}^K\vec{x}_k$ corresponding to $f^{(i)}$. Moreover, we use matrix notation and define vector $\vec{\bm{s}}$ as 
\begin{align}
\vec{\bm{s}}: = \bm{A} \Big(\mathds{1}_K\otimes \bm{x}\Big),    
\end{align}
where $\bm{A}\in \{0,1\}^{M\times qK}$  is a binary matrix that selects all the possible cases of nodes to send their value, and $\mathds{1}_K$ stands for a  vector of size $K\times 1$ whose all elements are one. 
To compute perfectly the function $f$ associated with modulated values  $\bm{x}$, we need to make sure that all $\vec{s}_i$'s do not have destructive overlapping and cover all the range of function $f$.  In particular, if $f^{(i)}$ is different from $f^{(j)}$, this imposes that the resulting constellation point $\vec{s}_i$ must not be the same as $\vec{s}_j$ for $i\neq j$ (see Fig. \ref{fig:range_vector}). To guarantee that digital modulation signals are suitable for computing the desired function $f$, we can formulate the following problem:
\vspace{-3pt}
\begin{subequations}
    \label{eq:feasibility}
\begin{align}
\nonumber
 \mathcal{P}_1 = {\rm find} &
& & ~~~~~~\bm{x} \\ 
{\rm s.t.}&
& & 
f^{(i)}\neq f^{(j)} \Rightarrow s_i \neq s_j,~ \forall (i,j) \in [M]^2,\\
& & &  \|\bm{x}\|_2= P,
\end{align}
\end{subequations}
where recall that $\bm{x} \in \mathbb{C}^{q\times 1}$ is the modulation vector, $[M]^2$ denotes $[M]\times[M]$, and $P$ is the power budget on the nodes. 

Problem $\mathcal{P}_1$ is a feasibility problem to check whether the constraints are feasible. Problem $\mathcal{P}_1$ is extremely difficult because the constraints are not convex and non-smooth. To overcome this difficulty, we replace them with smooth conditions as follows:
\begin{subequations}
    \label{eq:feasibility-convex}
\begin{align}
\nonumber
 \mathcal{P}_2\!=\!{\rm find} &
& & ~~~~~~\bm{x} \\
{\rm s.t.}&
& & \label{eq:feasibility-convex-a}
|(\bm{a}_i - \bm{a}_j)^{\mathsf{T}}\bm{x}|^2  \geq \gamma |f^{(i)} - f^{(j)}|^2, \forall (i,j) \in  [M]^2,\\
& & & \|\bm{x}\|_2= P,
\end{align}
\end{subequations}
where $\gamma>0$ is a positive normalization factor, and  $\bm{a}_i$ denotes $i$-th row of matrix $\bm{A}$. Note that for any small value of $\gamma$, the solution to Problem $\mathcal{P}_1$ is equal to the solution to Problem $\mathcal{P}_2$.

\begin{remark}
\label{rem:Power}
Problem  $\mathcal{P}_2$ not only satisfies the constraints of Problem  $\mathcal{P}_1$ but also designs the transmit constellation points for achieving the acceptable computation error in noisy communication channels. The reason is that the right side of the constraints is the computation error;  a higher computation error enforces a larger distance of constellation points and enforces more energy. In other words, the distances between constellation points are penalized based on possible computation errors. 
\end{remark}

Unfortunately, optimization problem \eqref{eq:feasibility-convex} is a quadratically constrained quadratic programming (QCQP) problem and known to be an \textit{NP-hard} \cite{Sidir2006Physical}. However, one can rewrite Problem $\mathcal{P}_2$ as a semidefinite programming problem and relax it using the \textit{lifting trick} \cite{vandenberghe1996semidefinite} by recasting the cost function in terms of the lifted variable $\bm{X}:= \bm{x}\bm{x}^{\mathsf{H}}$ as 
\begin{subequations}
    \label{eq:traceX-nonconvex}
\begin{align}
\nonumber
\mathcal{P}_3 = {\rm find} &
& & ~~\bm{X} \\
{\rm s.t.}&
& & 
 \langle \bm{X}, \bm{B}_{i,j} \rangle    \geq  {g}_{i,j},~  {\rm trace}(\bm{X}) = P, \\ \label{eq:rankcons}
& & & \bm{X} \succeq \bm{0}, \quad {\rm rank}(\bm{X}) = 1, 
\end{align}
\end{subequations}
where $ {g}_{i,j} = \gamma |f^{(i)}-f^{(j)}|^2$  for all $(i,j) \in [M]^2, i \neq j$ and $\bm{B}_{i,j} = (\bm{a}_i-\bm{a}_j)(\bm{a}_i-\bm{a}_j)^{\mathsf{T}}$. This is a linear and convex problem concerning matrix $\bm{X}$ except for the rank constraint in \eqref{eq:rankcons}. Next, we obtain the relaxed problem by dropping the rank-one constraint from the optimization \cite{Sidir2006Physical}, i.e.,
\begin{subequations}
    \label{eq:traceX-convex}
\begin{align}
\nonumber
\mathcal{P}_4=  {\rm find} &
& & \bm{X} \\
{\rm s.t.}&
& & 
 \langle \bm{X}, \bm{B}_{i,j}\rangle  \geq  {g}_{i,j},~ \\
& & & \bm{X} \succeq \bm{0},~{\rm trace}(\bm{X}) = P,
\end{align}
\end{subequations}
which can be solved by using the CVX solver \cite{grant2014cvx}. In the case where  the solution to the SDP problem of \eqref{eq:traceX-convex}, denoted as $\bm{X}^*$, is a rank-one matrix, we can obtain the optimal modulation vector $\bm{x}^*$ using the Cholesky decomposition of $\bm{X}^*$. Otherwise, we need to use the Gaussian randomization method \cite{luo2010semidefinite}, whose output is a sub-optimal solution to the original problem with a guaranteed optimality gap \cite{Miarton2021Full}.

We note that multiple antennas at the receiver and transmitter (MIMO) can lead to having a vector of values at the CP. Let us consider that $N_t$  and $N_r$  are the numbers of antennas at the transmitter and receiver, respectively. Then, we can compute a vector function $\bm{f}: \mathbb{R}^{N_t} \mapsto \mathbb{R}^{N_r}$, instead of only a single scalar as the output of the computation. Therefore, the MIMO extension of ChannelComp can be used for vector-based computation, such as in optimization theory, machine learning, and federated learning, in which one deals with a gradient vector instead of just a scalar.

\subsection{ChannelComp decoder}\label{sec:Reciever}
The decoder can be determined as long as the solution to Problem $\mathcal{P}_4$ satisfies its constraints. We use the maximum likelihood estimator (MLE) and design the decision boundaries based on the reshaped constellation points over the MAC. Next, the tabular function $\mathcal{T}(\cdot)$ maps the received signal $\vec{y}$ to the desired output of the function $f$.  Specifically, we define $\vec{g}_{i}\!:=\!\sum_{k=1}^K\vec{x}_k$ as the constellation point for the function $f^{(i)}$. Then, the problem is to find which $\vec{g}_{i}$'s values were transmitted while we have received $\vec{y}$. Hence, the MLE estimator gives us the following: 
\begin{align}
    \hat{f}^{(i)} = \underset{i}{\rm argmax}\Pr(\vec{y} | \vec{g}_{i}),    
\end{align}
where $\Pr(\vec{y} | \vec{g}_{i}) =  1/{\sqrt{2\pi\sigma_z^2}} \exp{\big[-\|\vec{y} -\vec{g}_i\|_2^2/2\sigma_z^2\big]}$ follows a Normal distribution. Next, taking logarithm results in: 
\begin{align}
\label{eq:funi}
     \hat{f}^{(i)}= \underset{i}{\rm argmin} \| \vec{y} - \vec{g}_i\|_2^2.    
\end{align}
Using the expression in \eqref{eq:funi},  the decoder generates the set of all possible constellation points $\{\vec{g}_1,\ldots,\vec{g}_M\}$ with the corresponding Voronoi cells $\{\mathcal{V}_1,\ldots, \mathcal{V}_M\}$. Then, the desired value is given by 
\begin{align}
    \hat{f} = \sum\nolimits_{j=1}^{M}\mathcal{T}_j(\vec{y}),
\end{align}
in which the look-up table is 
\begin{align}
      \mathcal{T}_j(\vec{g}):= \begin{cases}
            f^{(j)}, &    \vec{g}\in \mathcal{V}_j, \\
            0, & \text{ otherwise}.
        \end{cases}
\end{align}

For those cases where the computation conditions in \eqref{eq:feasibility-convex} are not  established, we have the same constellation point $\vec{g}_i$ at different values of the function, e.g., $f^{(i)}$ and $f^{(i+1)}$. Hence, these points have the same Voronoi cell, i.e., $\mathcal{V}_i = \mathcal{V}_{i+1}$, and the output values can be replaced by their mean, i.e.,
\begin{align}
 \mathcal{T}_i(\vec{g})= \mathcal{T}_{i+1}(\vec{g}) = \begin{cases}
            \frac{f^{(i)}+ f^{(i+1)}}{2},  &    \vec{g}\in \mathcal{V}_i, \mathcal{V}_{i+1}, \\
            0, & \text{otherwise.}
        \end{cases}
\end{align}
It is important to note that the encoder and decoder in ChannelComp have a similar overhead compared to AirComp. This is because we only map the input and output using the modulation vectors obtained from Problem $\mathcal{P}_4$. The main complexity comes from solving the optimization in \eqref{eq:traceX-convex}. This optimization must be done before setting up the communication system. 

In the following, we assess the performance of the proposed  encoder and decoder of ChannelComp over the MAC.

\begin{figure}
\centering
    \begin{tikzpicture} 
    \begin{axis}[
        label style={font=\scriptsize},
        xlabel={${\rm Real}(\bm{x})$},
        ylabel={${\rm Imag}(\bm{x})$},
        legend cell align={left},
        tick label style={font=\scriptsize}, 
        width=9cm,
        height=8cm,
        xmin=-5, xmax=7,
        ymin=-5, ymax=7,
        legend style={nodes={scale=0.8, transform shape}, at={(0.65,0.98)}}, 
        ymajorgrids=true,
        xmajorgrids=true,
        grid style=dashed,
        ticklabel style = {font=\scriptsize},
        grid=both,
        grid style={line width=.1pt, draw=gray!10},
        major grid style={line width=.2pt,draw=gray!30},
    ]
    \addplot[
        color=chestnut,
        mark=o,
        line width=1pt,
        mark size=2pt,
        ]
    table[x=X1,y=Y1]
    {Data/SumModulationK2Q3.dat};
       \addplot[
        color=airforceblue,
        mark=o,
        line width=1pt,
        mark size=2pt,
        ]
    table[x=X1,y=Y1]
    {Data/MaxK2Q3.dat};
       \addplot[
        color=cssgreen,
        mark=o,
        line width=1pt,
        mark size=2pt,
        ]
    table[x=X1,y=Y1]
    {Data/ProdModK2Q3.dat};
         \addplot[
        color=cadmiumorange,
        mark=o,
        line width=1pt,
        mark size=2pt,
        ]
    table[x=X,y=Y]
    {Data/SymQuadQ3.dat};
    \legend{$f(\bm{x}) = \sum_{k=1}^Kx_k$, $f(\bm{x}) = \max_{k}x_k$,$f(\bm{x}) = \prod_{k=1}^Kx_k$,$f(\bm{x}) = \sum_{k=1}^Kx_k^2$ };
    \end{axis}
\end{tikzpicture}
  \caption{ Constellation diagram of the modulation vector $\bm{x}$ for $q=8$ or $3$ bits which is the solution to the optimization in \eqref{eq:traceX-convex} for different functions. With these modulation vectors (which is the same for all nodes), in an AWGN free channel,  the constellation points diagram of $\vec{y}$ is uniquely mapped to the output values of the functions using a look-up table $\mathcal{T}(\vec{y})$. In the presence of AWGN, the constellation point of $\vec{y}$ will be centered around those of the AWGN free channel. There will be a resulting probability of computing error, which however  can be made as small as AirComp's one, or smaller.}
  \label{fig:ConsQ4K2}
   
\end{figure}
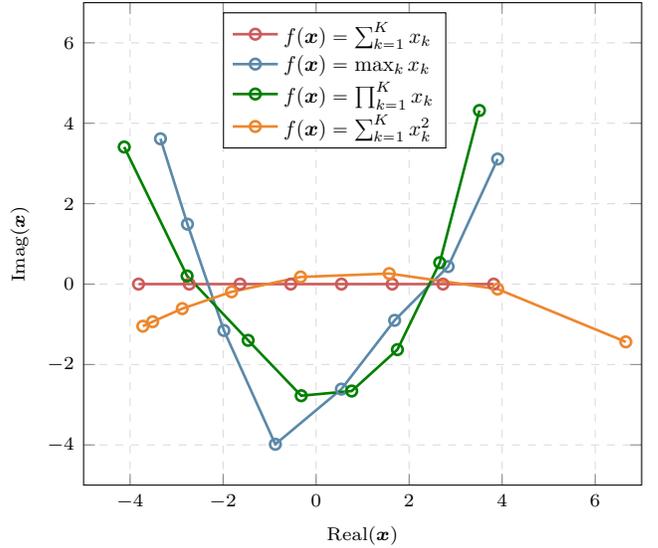
\section{Numerical Experiments}\label{sec:Num}
We evaluate the performance of Problem $\mathcal{P}_4$ for different functions and different numbers of nodes.
In the next subsection, we compare the performance between ChannelComp,  a standard digital transmission scheme, OFDMA, in which the modulation vector is obtained from \eqref{eq:traceX-convex}, and each node uses different frequency channels, and with AirComp \cite{Golden2013Harnessing}, which uses analog modulation. We compare these schemes in cases where the input values of function $f$ are continuous and discrete.



\subsection{Performance Evaluation of ChannelComp}
To evaluate the performance of ChannelComp, we solve the relaxed optimization problem in \eqref{eq:traceX-convex} where the channel is considered ideal and noiseless. This experiment is repeated for four functions, i.e., summation, product, maximum, and quadratic, which are symmetric functions, with a quantization level $q=8$  for each node. The resultant modulation vectors are depicted in Fig.  \ref{fig:ConsQ4K2} for all the functions. The resultant digital modulation vectors satisfy all the constraints in Problem $\mathcal{P}_2$, which means they do not cause any confusing  overlapping over-the-air. Hence, ChannelComp makes it possible to have an error-free computation. For the summation function, the output of ChannelComp is consistent with that of the PAM modulation. This is a result of the observation made in Remark \ref{rem:Power}. Indeed, one may think of using QAM modulation instead to be more power efficient, leading to more computation errors. 

\subsection{Comparsion to Analog AirComp}

 In the next experiment, the \Comp is compared to analog AirComp for the summation and product functions\footnote{
For other functions, such as the maximum function, i.e., $ f = \max_k x_k$, AirComp approximates the maximum using the log sum function \cite{sahin2022survey}, which is an approximation and is not the exact value even in a noise-free MAC. However, in ChannelComp, we are able to compute the exact value.}.

To do this, we consider two scenarios. For the first scenario, we consider that the input signal has discrete values, thus the quantization module is not necessary. Therefore, we transmit the modulated value over the AWGN channel. Specifically, we consider functions $f_1\!=\!\sum_{k=1}^{4}x_k$ and $f_1\!=\!\prod_{k=1}^{4}x_k$, where $x_k\in \{0,1,2, \ldots, 7\}$ over a network with $K=4$ nodes.  Note that in OFDMA, every node uses the same modulation as in ChannelComp, but with different frequencies. The modulation vector is determined through the optimization problem \eqref{eq:traceX-convex}. 
The normalized mean square error (NMSE) metric is used for characterizing the computation error, which is defined as 
${\rm NMSE}:= \sum_{j=1}^{N_s}|f^{(i)} - \hat{f}_j^{(i)}|^2/{N_s|f^{(i)}|}$,  where $f^{(i)}$ denotes the value of the desired function, and $\hat{f}_j^{(i)}$ is the $j$-th estimated value of $f^{(j)}$ for $j\in [N_s]$, and $N_s$  denotes the number of Monto Carlo trials. 

Fig. \ref{fig:QunNMSE} shows the different signal-to-noise ratios (SNRs), which is defined as ${\rm SNR}:=20\log(\|\bm{x}\|_2/\sigma_z)$. Note that the performance of the ChannelComp method in computing the summation function outperforms AirComp, particularly for the high SNR region.  For the product function, the performance of the AirComp and OFDMA decreases by $10$ dB while the \Comp can obtain similar performance with computing the summation function due to the reshaped modulation points to be fitted with the output of function $f$. Moreover, when the noise variance is very high (SNR less than $-3$ dB), OFDMA performs better than ChannelComp in the summation function, which comes from the input values having limited domain\footnote{When the variance of the noise is very high, we only observe the value of the boundary of the input domain with high probability. In fact, for OFDMA, with high probability, the estimation is either $0$ or $7$. As a result, the estimation of the summation is more likely around $14$ for $K = 4$ nodes. However, for ChannelComp, we directly compute the summation. Accordingly, the estimation of the summation would be either $0$ or $28$ with high probability, which leads to more error. } ($x_k$s are between $0$ and $7$).  
\begin{figure}
    \begin{tikzpicture} 
    \begin{axis}[
        xlabel={SNR (dB)},
        ylabel={NMSE},
        label style={font=\scriptsize},
        legend cell align={left},
        tick label style={font=\scriptsize} , 
        width=8cm,
        height=7cm,
        xmin=-5, xmax=27,
        ymin=1e-4, ymax=1e2,
        ymode = log,
       legend style={nodes={scale=0.75, transform shape}, at={(0.98,0.98)}}, 
        ymajorgrids=true,
        xmajorgrids=true,
        grid style=dashed,
        grid=both,
        grid style={line width=.1pt, draw=gray!15},
        major grid style={line width=.2pt,draw=gray!40},
    ]
    \addplot[
        color=bleudefrance,
        mark=o,
        line width=1pt,
        mark size=2pt,
        ]
    table[x=SNR,y=ChannelComp]
    {Data/MSEQError.dat};
     \addplot[
        color=cadmiumorange,
        mark=o,
        line width=1pt,
        mark size=2pt,
        ]
    table[x=SNR,y=OAT]
    {Data/MSEQError.dat};
    \addplot[
        color=cadmiumgreen,
        mark=o,
        line width=1pt,
        mark size=2pt,
        ]
    table[x=SNR,y=OFDM]
    {Data/MSEQError.dat};
    \addplot[
        color=bleudefrance,
        mark=triangle,
        line width=1pt,
        mark size=2pt,
        ]
    table[x=SNR,y=ChannelComp]
    {Data/MSEProdcutError.dat};
     \addplot[
        color=cadmiumorange,
         mark=triangle,
        line width=1pt,
        mark size=2pt,
        ]
    table[x=SNR,y=OAT]
    {Data/MSEProdcutError.dat};
    \addplot[
        color=cadmiumgreen,
        mark=triangle,
        line width=1pt,
        mark size=2pt,
        ]
    table[x=SNR,y=OFDM]
    {Data/MSEProdcutError.dat};
    \legend{ChannelComp $\sum $,AirComp$\sum $, OFDMA$\sum $,ChannelComp $\prod $,AirComp$\prod $, OFDMA$\prod $};
    \end{axis}
\end{tikzpicture}

  \caption{Performance comparison between our proposed ChannelComp, the traditional AirComp, and OFDMA methods in terms of NMSE error averaged over $N_s =100$ when values of the function to be computed are originally quantized. The input values are $x_k=\{0, 1, 2, \ldots, 7\}$  and the desired functions are $f_1 = \sum_{k=1}^4x_k$ and $f_2 = \prod_{k=1}^4x_k$. }
  \label{fig:QunNMSE}
   
\end{figure}
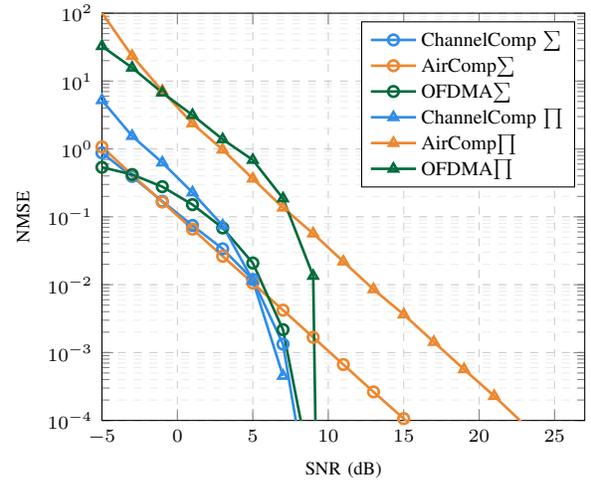
\begin{figure}
    \begin{tikzpicture} 
    \begin{axis}[
        xlabel={SNR~(dB)},
        ylabel={NMSE},
        label style={font=\scriptsize},
        legend cell align={left},
        tick label style={font=\scriptsize} , 
        width=8cm,
        height=7cm,
        xmin=-5, xmax=27,
        ymin=1e-3, ymax=1.5,
        ymode = log,
       legend style={nodes={scale=0.75, transform shape}, at={(0.98,0.98)}}, 
        ymajorgrids=true,
        xmajorgrids=true,
        grid style=dashed,
        grid=both,
        grid style={line width=.1pt, draw=gray!15},
        major grid style={line width=.2pt,draw=gray!40},
    ]
    \addplot[
        color=bleudefrance,
        mark=triangle,
        line width=1pt,
        mark size=2pt,
        ]
    table[x=SNR,y=ChannelCompQ1]
    {Data/MSEQWError.dat};
    \addplot[
        color=bleudefrance,
        mark=triangle,
        mark options = {rotate = 180},
        line width=1pt,
        mark size=2pt,
        ]
    table[x=SNR,y=ChannelCompQ2]
    {Data/MSEQWError.dat};
  \addplot[
        color=cadmiumorange,
        dashed,
        line width=1pt,
        ]
    table[x=SNR,y=OAT]
    {Data/MSEQWError.dat};
     \addplot[
        color=cadmiumgreen,
        mark=triangle,
        line width=1pt,
        mark size=2pt,
        ]
    table[x=SNR,y=OFDMQ1]
    {Data/MSEQWError.dat};
     \addplot[
    color=cadmiumgreen,
    mark=triangle,
    mark options = {rotate = 180},
    line width=1pt,
    mark size=2pt,
    ]
table[x=SNR,y=OFDMQ2]
{Data/MSEQWError.dat};
    \legend{ChannelComp $q =4$,ChannelComp $q = 16$,AirComp, OFDMA $q=4$,OFDMA $q=16$};
    \end{axis}
\end{tikzpicture}

  \caption{Performance comparison between ChannelComp, AirComp, and OFDMA in terms of NMSE error for computing the summation function  $f = \sum_{k=1}^4x_k$ with continuous uniform random input values $x_k$ in the interval $[0,7]$, averaged over $N_s =100$ Monto Carlo trials. }
  \label{fig:QunWithNMSE}
   
\end{figure}
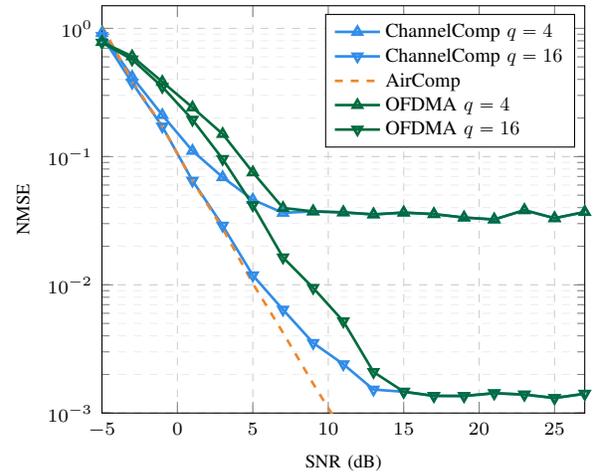
In the second scenario (see Fig. \ref{fig:QunWithNMSE}), we generate continuous uniform random numbers between $0$ and $7$, i.e., $x_k \sim [0,7]$ for computing the summation function, which has shown better performance for AirComp in the previous experiment. Afterward, for the ChannelComp, these values are quantized with $q=4$ and $q=16$ levels ($2$ and $4$ bits) and transmitted over the MAC channel. As can be observed from Fig. \ref{fig:QunWithNMSE}, the performance of \Comp is saturated by the quantization noise level when SNR is low. However,  increasing the number of bits can mitigate this issue. We can observe that \Comp outperforms both the OFDMA and the AirComp methods while using $1/K\times$ fewer communication resources (e.g., bandwidth or time) than the OFDMA method.

\section{Conclusion}\label{sec:conclusion}

 We proposed a fundamentally new over-the-air computation principle and method that uses digital modulations to compute functions over multiple access channels. We called the proposed method ChannelComp. We showed that it can compute a much more general class of functions than the well-known AirComp method, a method that is restricted to analog modulations. Moreover, similar to analog AirComp, ChannelComp can handle massive devices simultaneously so that the computation time can be strictly constrained.
 
 We proposed an encoder method based on a feasibility function's optimization problem, and a decoder that can compute functions using digital modulations. Finally, the simulation results showed that \Comp overall outperforms analog AirComp and OFDMA methods in terms of normalized mean squared error with around $10$ dB improvement in terms of computation error.

 There are a plethora of  interesting potential extensions of \Comp that we are planning to investigate.  In the future, we will extend \Comp for general functions using different modulations for each node and evaluate the effect of stochastic fading. Moreover, the current single narrowband antenna at the transmitters and the receiver system model can be extended to broadband multiple inputs and outputs to be able to compute vector-based calculations for applications such as federated learning. Furthermore, we will show that ChannelComp can significantly enable  applications such as federated edge learning.

\section{Acknowledgment}

This work is partially supported by the Wallenberg AI, Autonomous Systems and Software Program (WASP), the European Union’s Horizon Europe research and innovation program under the Marie Skłodowska-Curie project FLASH, with
grant agreement No 101067652. Also, the EU FLASH project,
the Digital Futures project DEMOCRITUS, and the Swedish
Research Council Project MALEN partially supported this
work.



\bibliographystyle{IEEEtran}
\bibliography{IEEEabrv,Ref2}

\end{document}